\documentstyle[12pt,epsf]{article}

\begin{document}
\begin{center} {\Large {\bf What is the Reference Frame of
an Accelerated Observer?
\footnote{Phys. Lett. A {\bf 215}, p.1 (1996).}}}\\[1cm]
Karl-Peter Marzlin
\footnote{e-mail: peter.marzlin@uni-konstanz.de} \\[2mm]
Fakult\"at f\"ur Physik
der Universit\"at Konstanz\\
Postfach 5560 M 674\\
D-78434 Konstanz, Germany
\end{center}
$ $\\[3mm]
\begin{minipage}{15cm}
\begin{abstract}
The general construction of extended reference frames for
non-inertial observers in flat space is studied. It is shown
that, if the observer moves inertially before and
after an arbitrary acceleration and rotation, the region where
reference frames can coincide with an inertial system is
bounded for final velocities exceeding 0.6 $c$.
\end{abstract}
\end{minipage}
$ $ \\[1cm]
\newpage
\section{Introduction}
One of the most basic concepts in modern physics is that
of an inertial system, i.e., the extended reference frame of an
observer with constant velocity in flat space.
This is not only a keystone in special relativity. Also
Einstein's derivation of the general theory of relativity was
inspired by the idea to extend the equivalence of the
inertial frames in Special Relativity
to an equivalence of all frames, including
those of accelerated observers. It is therefore of interest to ask
how such a frame may be realized in general relativity.

A second and more important reason for the study of extended
reference frames in general relativity is Mashhoon's discovery
of the problem with the
hypothesis of locality which is needed for the principle of
general covariance \cite{mash86}.
This problem is based on the fact that an
acceleration $a$ introduces a length scale $c^2 /a$. Mashhoon
argued that
the hypothesis of locality, i.e., the local equivalence between
an accelerated observer and an inertial observer having the same
velocity at a given space-time point,
holds only for wavelengths much smaller than this scale.
Later \cite{mash90} he demonstrated
this problem with an example based on Fermi coordinates.
It is the working hypothesis of the present paper that it may
be possible to overcome the restriction to small wavelengths by
introducing an extended reference frame for accelerated observers
in which the coordinates of an event in space-time
are identified with the proper time
and certain proper lengths between the worldline and the event
even if it is far away from the worldline.
An extended frame of reference may circumvent the problem
because it was (possibly) constructed by use of a non-local
procedure. The price one has to pay for this solution
is that the equivalence of all coordinate systems is lost for the
observer who obviously distinguishes the extended frame of
reference from other systems. The principle that all
extended coordinate
systems are equally valid is not affected, however,
as long as we do not consider a particular observer.

A very good approximation to the notion of a
frame of reference in the context of general relativity are
Fermi coordinates \cite{manasse63,ni78}. In this coordinate
system the metric of space-time is Minkowskian on the complete
world line of the observer. In the limit of weak
gravitational fields the Newtonian potential can be recovered
in this coordinate system \cite{kpm94}. An important point is
that the values of the spatial coordinates $x^i$
are directly related to the
proper length $s$ of geodesics connecting the observer with
other space-time points by $ [(x^1)^2 +(x^2)^2 + (x^3)^2 ]^{1/2}
= s$. They are therefore measurable quantities. The success of
Fermi coordinates in the neighborhood of the observer's worldline
may justify the following criterion for a frame of reference:
{\em A necessary condition for a coordinate system to be the
reference frame of a particular observer is that the metric in this
coordinate system is Minkowskian on the whole worldline}. The
physical content of this criterion is that locally the time and
lengths measured by the observer agree with the coordinates as
is globally the case in an inertial system. It clarifies the
notion of a reference frame as it is used in this paper.

Despite their
advantages Fermi coordinates have a severe shortcoming: they
cannot cover the complete space-time or even the past of
the observer if an acceleration is present.
An instructive example for this to occur is
Rindler space which is identical to the Fermi coordinate system
of an observer with constant acceleration in two dimensional
Minkowski space. It is well known that Rindler space
encloses only the right wedge of Minkowski space.
The rest of Minkowski space
cannot be covered by this coordinate system.

To circumvent these problems a modification of the original
construction scheme for Fermi coordinates was proposed
in Ref.~\cite{kpm94b}. In an ordinary Fermi
coordinate system the spatial coordinate lines are identified
with geodesics emanating from the world line and being
orthogonal
to the four-velocity of the observer. In the modified approach
these curves are not geodesics but are found
by a non-local construction scheme which guarantees that the
coordinates are directly related to the proper length of these
curves. In this point of view {\em the problem to find a
reference frame for an arbitrary observer is essentially
identical to the problem of choosing a sequence of spacelike
hypersurfaces belonging to this observer and defining his
slices of constant proper time}. Fermi coordinates are
a means to find at least a local approximation to these
planes of simultaneity; a modification of Fermi coordinates
should have the goal to construct the complete hypersurfaces
of constant proper time. It should be remarked that
the construction of an adapted coordinate system
is not the only way to deal with non-inertial observers.
Mashhoon \cite{mash93} has made a non-local ansatz
to overcome the problems with the hypothesis of locality, for
instance.

The central result of this paper will be a statement about the
general behaviour of any possible extended
reference frame, regardless how
it may be constructed. This result holds for observers in flat
space with
arbitrary worldlines and with any rotation provided both
acceleration and rotation will be non-zero only during a finite
period of proper time $T$. The study of this behaviour was
inspired by the examination of the particular proposal for an
extended frame of reference made in Ref. \cite{kpm94b}. For
illustration this special case is added in the last section
of this paper.
\section{Where can reference frames be Minkowskian?}
In the study of the modified Fermi coordinates introduced in
Ref. \cite{kpm94b} it turned out that for an observer constantly
accelerated during a finite period of time $T$ the coordinate lines
of proper time $\tau$ can become spacelike far away
from the worldline $z^\mu (\tau)$.
This will be briefly described in the next section.
Here it will be argued that this phenomenon occurs for a very
broad class of possible extended reference frames.

To demonstrate this we consider an observer which is in inertial
motion and without rotation
before the proper time $\tau =0$ and after
$\tau =T$. For simplicity we will first consider the two dimensional
case, in which a rotation is excluded, and switch to an inertial system in
which the observer is at
rest before $\tau =0$ and where $z^\mu(\tau =0)$ lies in the origin.
Attached to the worldline is a tetrad $e^\mu_{\underline{a}}$
with $e^\mu_{\underline{0}}$ being the tangent to the worldline and
the spatial tetrads $e^\mu_{\underline{i}}$ being orthogonal to it.
Throughout the paper greek indices run from 0 to 3, latin indices
run from 1 to 3, and underlined indices are tetrad indices.
The spatial coordinate lines start on the worldline with their
directions being given by the spatial tetrads (as for Fermi
coordinates). This guarantees that the metric is Minkowskian on the
whole worldline.

Without referring in any way to the actual construction
of the reference frame one can make two assumptions on how
it may look like.\\
1) As long as the observer is at rest its frame of reference
should certainly be an inertial system
at every point with causal connection to the world line.
This is the case for any point inside the lower light cone
in Fig.~1. \\
2) After the observer has reached his final velocity and is
not accelerated anymore the frame of reference should again be
an inertial system in a certain region around the worldline.
The central question of this paper is how large this region may
be.

Since the coordinate lines of an inertial system in flat space are
straight lines one can infer that (in two dimensions) the
coordinate transformation to the reference frame in the inertial
regions takes the form
\begin{eqnarray}
x^\mu(\tau,r) &=& z^\mu(T) + (\tau -T)
  e^\mu_{\underline{0}}(T) + r e^\mu_{\underline{1}}(T)
  \mbox{ for } \tau > T \nonumber \\
x^\mu(\tau,r) &=& z^\mu(0) + \tau
  e^\mu_{\underline{0}}(0) + r e^\mu_{\underline{1}}(0)
  \mbox{ for } \tau <0
\label{coord} \end{eqnarray}
(we use the convention sgn$(g_{\mu \nu}) = +2$).
In these equations $r$ denotes the proper distance between the
point $z^\mu(\tau)$ on the worldline and the point $x^\mu (\tau,r)$
(inside an inertial region)
on the spatial coordinate line that is parametrized by $r$. Hence
two points $x^\mu(\tau,r)$ and $x^\mu(\tau^\prime,r)$
having the same
parameter $r$ belong to the same line of constant proper distance
$r$ even if they lie in different inertial regions (compare Fig.
1). These lines are just the coordinate lines of proper time.

It is now reasonable to {\em demand that the coordinate lines of
the proper time are timelike everywhere}. A necessary condition
for this to hold is that the distance between two points having
the same parameter $r$ is always timelike. To check if this
condition is always fulfilled we choose two points.
One point $x^\mu(\tau = r,
r)$ is placed on the past light cone of the origin (units with
$c=1$ are used and $r<0$ is assumed) and therefore lies on the
borderline of the lower inertial region.
The other point $x^\mu(\tau^\prime ,r)$
belongs to the outgoing inertial region (compare
Fig. 1). In the inertial system
that was chosen the various tetrad vectors in Eq. (\ref{coord})
can be represented as $e^\mu_{\underline{0}}(0) =(1,0)$,
$e^\mu_{\underline{1}}(0) =(0,1)$, $e^\mu_{\underline{0}}(T) =
(\cosh \alpha,\sinh \alpha)$, and $e^\mu_{\underline{1}}(T) =
(\sinh \alpha,\cosh \alpha)$ so that $\Delta x^\mu :=
x^\mu (\tau^\prime,r) - x^\mu(\tau=r,r)$ can be calculated
explicitly. It is clear that $\Delta x^\mu$ is a timelike vector
if $r$ is small. If it becomes spacelike for growing $|r|$ there
must be a value $\hat{r}$ of $r$ for which it is lightlike. To
determine $\hat{r}$ one has to solve the equation $\Delta x^\mu
\Delta x_\mu = 0$ or, in two dimensions and since both $x^\mu (
\tau^\prime,r)$ and $x^\mu (\tau=r,r)$ are on the left of the
(timelike) worldline, $\Delta x^0 + \Delta x^1 =0$. Inserting
Eq. (\ref{coord}) into this expression leads to
\begin{equation}
d e^\beta + (\tau^\prime -T) e^\alpha + \hat{r}(e^\alpha -2) =0
\label{dxmu} \end{equation}
where the timelike vector $z^\mu(T) -z^\mu(0)$ was written as $(d
\cosh \beta , d \sinh \beta)$ with $d$ being positive.

The first two terms on the l.h.s. of Eq. (\ref{dxmu}) are always
positive so that the last must be negative in order to fulfill the
equality. Since $r<0$ this can
only happen for $\exp (\alpha) >2 $. This implies that the final
velocity $v$ of the observer must be larger than
\begin{equation}
v = \tanh (\alpha) > 0.6 \mbox{ for } e^\alpha >2 \; .
\end{equation}
For velocities larger than 0.6 one can determine $\tau^\prime$
as function of $\hat{r}$ (or vice versa) from Eq. (\ref{dxmu}).
This results in a borderline $x^\mu(\tau^\prime (\hat{r}), \hat{r})$
above which a reference frame is
allowed to behave as an inertial system without necessarily
leading to a spacelike time coordinate. It is given by
\begin{equation}
x^\mu(\tau^\prime (\hat{r}), \hat{r}) = z^\mu(T) - d e^{\beta -
  \alpha} e^\mu_{\underline{0}}(T) + \hat{r} \big \{ (2 e^{-\alpha}
  -1) e^\mu_{\underline{0}}(T) - e^\mu_{\underline{1}}(T) \big \}
\; .\label{border} \end{equation}
This is a spacelike straight line approaching the light cone if
the velocity $v$ is close to 1. The line starts at the point
$z^\mu(T) - d e^{\beta - \alpha} e^\mu_{\underline{0}}(T)$. It is
not difficult to show that this point is given by the intersection
point of the future light cone of the origin and a hypothetical
inertial observer's  worldline which agrees with $z^\mu(\tau)$
for $\tau >T$.

Putting everything together we found that {\em if in any construction
of a reference frame the coordinate lines of proper time are
always timelike then for a final velocity $v>0.6$
the inertial system of the outgoing observer must lie above the
spacelike straight line given by Eq. (\ref{border}). This line
approaches the light cone if $v$ approaches the velocity of light.}

This result may be generalized in several ways. The first question
is if the result also holds in four dimensions and in the presence of
a rotation. This can be answered
by working in an inertial system where the observer
is initially at rest in the origin and where the final velocity
points towards the $x^1$ direction. If only coordinate lines
are studied which for $\tau <0$ are in the $x^0-x^1$ plane then
for vanishing rotation the situation is the same as in the two
dimensional example shown in Fig. 1 and Eq. (\ref{coord}). The
new feature in four dimensional space is that in case of a
rotated observer the outgoing coordinate line could be rotated,
too. Hence this line need not to lie in the $x^0-x^1$ plane for
$\tau > T$.

More formally, if one introduces in the outgoing region
a second set of four orthogonal vectors,
\begin{eqnarray}
  \bar{e}^\mu_{\underline{0}} = (\cosh \alpha , \sinh \alpha, 0,0)
  &,& \bar{e}^\mu_{\underline{1}} = (\sinh \alpha , \cosh \alpha,
  0,0) \nonumber \\
  \bar{e}^\mu_{\underline{2}} = (0 ,0 , 1,0) &,&
  \bar{e}^\mu_{\underline{3}} = (0 ,0 , 0, 1)\; ,
\end{eqnarray}
the relevant tetrad vectors of the observer can be expressed as
$e^\mu_{\underline{0}}(T) = \bar{e}^\mu_{\underline{0}}$ and
$e^\mu_{\underline{1}}(T) = c_1 \bar{e}^\mu_{\underline{1}} +
c_2 \bar{e}^\mu_{\underline{2}} +c_3 \bar{e}^\mu_{\underline{3}}$
with $c_1^2 +c_2^2+c_3^2 =1$. In this notation $\Delta x^\mu$ is
given by
\begin{equation}
 \Delta x^\mu = z^\mu(T) +(\tau^\prime -T) e^\mu_{\underline{0}}(T)
  + r \big \{ c_1 \bar{e}^\mu_{\underline{1}} +
  c_2 \bar{e}^\mu_{\underline{2}} +c_3 \bar{e}^\mu_{\underline{3}}
  - e^\mu_{\underline{0}}(0)-e^\mu_{\underline{1}}(0)\big \}
  \; . \label{4dim}\end{equation}
To avoid tedious calculations it is convenient to calculate
$\Delta x^\mu \Delta x_{\mu}$ only for the case $r$ and $\tau
^\prime$ being large. Then the term
$z^\mu (T)$ in Eq. (\ref{4dim}) is negligible and one finds
\begin{equation} \Delta x^\mu \Delta x_{\mu} = -(\tau^\prime -T)^2
  + r^2 [1-2 c_1 e^{-\alpha}] + 2 r (\tau^\prime -T) e^{-\alpha} \;
  .\end{equation}
This expression agrees with the two dimensional result for $c_1=1$
if $z^\mu(T)$ is also neglected in the two dimensional case.
For $c_1 < 1$ the expression for
$\Delta x^\mu \Delta x_{\mu}$ becomes larger, that is "more
spacelike". Thus for long times $\tau^\prime$ a rotation will
lead to even more stringent conditions on the inertial regions
than those found in two dimensions.
{\em The result presented for the two dimensional case is therefore
also a lower bound for the four dimensional case which includes
a possible rotation of the observer}.
\section{A particular model}
To demonstrate the result presented above for a particular
construction of a possible reference frame the proposal of
Ref. \cite{kpm94b} will be used. In this model
the hypersurfaces of constant proper time
are constructed according to the following principle:
(the tangents of) the spatial coordinate
lines should, after parallel transport along a light ray, always
be orthogonal to the four-velocity $dz^\mu /d \tau$
of the observer at the point in the future of the coordinate line
where the light ray intersects the world line. All details can
be found in Ref. \cite{kpm94b}. The motivation to study this
proposal was to get rid of the shortcomings of the original
construction for Fermi coordinates.
In the present context it is essential
that the spatial coordinate lines are
constructed by using only the past light cone of the points on
the worldline, not their future light cones. Therefore any
point on the spatial
coordinate lines contains only information about that part of the
worldline which lies in its future. This implies that after the
observer reached the final
velocity all coordinate lines are constructed only with the
aid of an inertial worldline and hence are straight lines.

Without going into any details we state here the results for
an observer who is accelerated with constant acceleration $a$
between $\tau =0$ and $\tau =T$. In this case the
differentiable worldline is given by
\begin{equation} \left ( \begin{array}{c} z^0 (\tau) \\ z^1
     (\tau) \end{array} \right ) = \left \{ \begin{array}{cl}
     \left ( \begin{array}{c} \tau \\ 1/a
     \end{array} \right ) & \tau < 0 \\
     \left ( \begin{array}{c} \sinh (a \tau)/a \\ \cosh (a \tau
     )/a \end{array} \right ) & 0 < \tau < T \\
     \left ( \begin{array}{c} \sinh (a T)/a + (\tau -T) \cosh
     (aT) \\ \cosh (a T)/a +(\tau -T) \sinh (a T)
     \end{array} \right ) & \tau > T \end{array} \right .
     \label{worldline} \end{equation}
It should be noted that for this worldline the spatial coordinate
lines of the original Fermi coordinates do intersect far away
from the worldline so that the time coordinate becomes undefined.
This also shows that ordinary Fermi coordinates can only be
considered as a local approximation to an extended reference
frame. Following the steps of Ref. \cite{kpm94b} the
metric in the new "reference frame" is given by
\begin{equation} g_{\mu \nu} = \left (\begin{array}{cc} 1-2\exp
     (a\phi ) & \mbox{sgn} (r)\, [1- \exp (a\phi )] \\
     \mbox{sgn} (r)\, [1- \exp (a\phi )] & 1 \end{array}
     \right ) \label{metendacc}\end{equation}
where the function $\phi$ is defined by
\begin{equation} \phi (\tau ,r) := \Theta (T-\tau)\,
     S(r, T-\tau) -
     \Theta (-\tau)\, S(r, -\tau) \end{equation}
with $\Theta$ being the step function and
\begin{equation} S(x,q) := \left \{ \begin{array}{cl} q &
     ,x>q \\ x & ,-q < x < q \\ -q &
     ,x<-q \end{array} \right . \end{equation}
The coordinate lines of this coordinate system for $v = \tanh (a T)
> 0.6$ are shown in  Fig. 2. One can see that the crossing of
spatial coordinate lines, which occurs in the original Fermi
coordinates, is avoided. The price we have to pay for this is
that the time coordinate
becomes spacelike far away from the worldline.

It is a pleasure for me to thank Ulf Jasper,
Claus L\"ammerzahl, and
Rainer M\"uller for stimulating
discussions and the Studienstiftung des Deutschen Volkes
for financial support.

\newpage
\begin{figure}[t]
\epsfysize=7cm
\hspace{3cm}
\epsffile{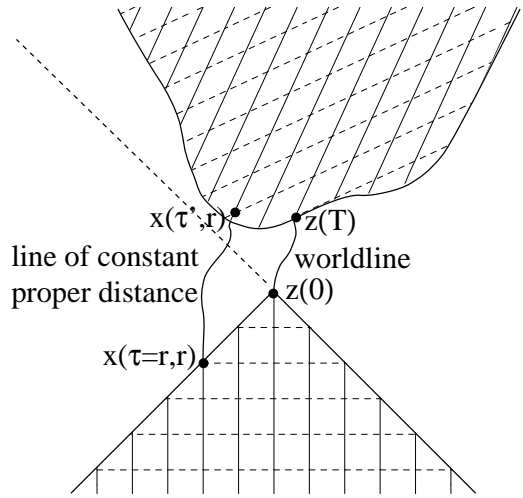}
\caption{The incoming and outgoing inertial regions of a possible
reference frame. Lines of constant $r$ are drawn solid,
and lines of constant $\tau$ are drawn dashed.
The thick solid line represents the world line. The dashed
line crossing the line of constant proper distance $r$ is the
boundary below which the reference frame cannot be inertial.}
\end{figure}

$ $

$ $

\begin{figure}[t]
\epsfysize=7cm
\hspace{3cm}
\epsffile{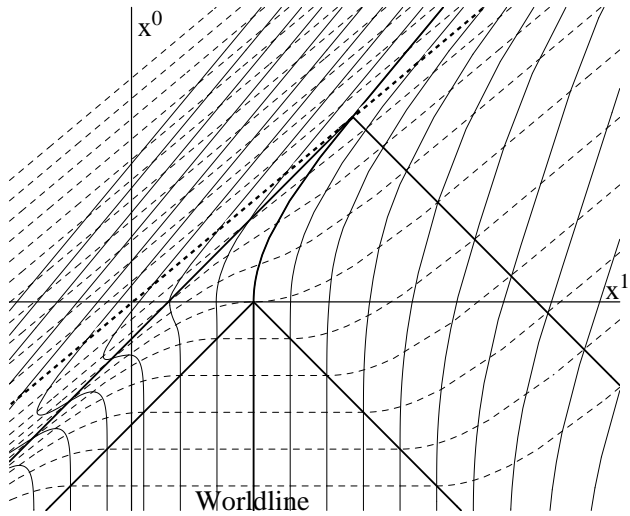}
\caption{The coordinate lines of the modified Fermi coordinates
for a final velocity larger than 0.6 $c$.
In this case the lines of constant
spatial coordinate $r$ can become spacelike.
The hypersurface of
constant proper time $\tau=T$ is drawn as a thick dashed line.}
\end{figure}

\end{document}